\documentclass[useAMS,usenatbib]{mn2e}
\usepackage{graphicx}
\usepackage{fixltx2e}
\title[On reconnection and topology in solar flux emergence]{On magnetic reconnection and flux rope topology in solar flux emergence}
\author[D. MacTaggart and A.L. Haynes]{D. MacTaggart$^{1}$\thanks{E-mail:
d.mactaggart@abertay.ac.uk (DM); andrew@mcs.st-and.ac.uk (ALH)}  and A.L. Haynes$^{2}$\\
$^{1}$White Space Research, University of Abertay Dundee, Kydd Building, Dundee, DD1 1HG, UK\\
$^{2}$School of Mathematics and Statistics, University of St Andrews, St Andrews, Fife, KY16 9SS, UK}
\begin{document}

\date{}

\pagerange{\pageref{firstpage}--\pageref{lastpage}} \pubyear{2013}

\maketitle

\label{firstpage}

\begin{abstract}
We present an analysis of the formation of atmospheric flux ropes in a magnetohydrodynamic (MHD) solar flux emergence simulation. The simulation domain ranges from the top of the solar interior to the low corona. A twisted magnetic flux tube emerges from the solar interior and into the atmosphere where it interacts with the ambient magnetic field. By studying the connectivity of the evolving magnetic field, we are able to better understand the process of flux rope formation in the solar atmosphere. In the simulation, two flux ropes are produced as a result of flux emergence. Each has a different evolution resulting in different topological structures. These are determined by plasma flows and magnetic reconnection. As the flux rope is the basic structure of the coronal mass ejection (CME), we discuss the implications of our findings for solar eruptions. 
\end{abstract}


\section{Introduction}
Coronal mass ejections (CMEs) are the most violent eruptions in the solar system. They can eject 10$^{13}$kg of plasma at 1000km/s into interplanetary space \citep{chen11}. Spectacular images from the \emph{Solar Dynamics Observatory} (\emph{SDO}) reveal near-circular loop structures that carry plasma from the Sun into space \citep[e.g.][]{koleva12}. As CMEs are responsible for some of the more destructive aspects of space weather, it is important to understand all aspects of their evolution, from formation to eruption. There exists a variety of models whose aims are to describe the different aspects of the CME life-cycle. Here we shall list some that focus on the evolution of a flux rope. All of these model the solar magnetic field using the magnetohydrodynamic (MHD) approximation. Some models start with an unstable twisted flux tube (hereafter called a flux rope) placed in the model corona. Via different ideal MHD instabilities, e.g. the kink and torus instabilities \citep{bateman78,hood79}, the flux ropes can expand rapidly into the corona and achieve typical CME speeds \citep{torok05,torok07}.

The main dynamical drivers of the solar atmosphere are emerging active regions. As these are the nurseries of CMEs, the relationship between active region evolution and CME formation/eruption is an important topic. There exists a series of numerical models that assume the presence of an active region magnetic field in the initial condition. This is a field with a simple topology, typically an arcade or a potential bipolar region. The field is then deformed by the imposition of motions on the lower boundary of the computational domain (normally taken to model the photosphere). Depending on the model, these motions are shears \citep{kusano05,aulanier12}, rotations \citep{torok05} and/or compressions \citep{amari03}. The common result is that an atmospheric flux rope is produced from the deformed active region field through magnetic reconnection. The resulting flux rope can then erupt as a CME through various mechanisms. For example, \citet{aulanier10} identify the eruption mechanism in their model to be the torus instability. In the breakout model \citep{antiochos99}, reconnection occurs above and below the flux rope. The reconnection above the rope weakens the tension of the overlying magnetic field, clearing the path ahead of it. The reconnection below the rope replenishes its flux and pushes it upwards.

Another class of models includes the effects of active region emergence. These can be divided into kinematic and dynamic. In kinematic models, flux ropes emerge through the lower boundary of the computational domain (photosphere) by imposed motions. These ropes interact with the coronal magnetic field and, if the conditions are suitable, can become unstable and erupt as CMEs. For example, \citet{fan07} drive a flux rope, quasi-statically, into their domain and find that it eventually erupts due to the torus instability.

In dynamic models of flux emergence, the computational domain normally models a region from the top of the solar interior to the low corona. A magnetic field (e.g. a flux rope or flux sheet) is placed in the solar interior and is either made buoyant or is given an initial upward velocity. It is then left to evolve self-consistently. There exists a large number of dynamic flux emergence simulations \citep[e.g.][]{magara03,manchester04,arber07,dmac09b,fan09,dmac11,fang12}. From these, and others, a general picture of flux rope emergence has developed. The magnetic field rises to the photosphere where it cannot continue to rise due to buoyancy alone. Here, a magnetic buoyancy instability \citep{acheson79,paris84} occurs, allowing the field to reach the corona. More details can be found in a recent review by \citet{hood12}.

In relation to the self-consistent formation of atmospheric flux ropes (candidates for CMEs) within emerging regions, there have been several recent dynamic flux emergence (hereafter referred to just as flux emergence) simulations that address this. \citet{archontis08} simulate the emergence of a twisted magnetic cylinder that is placed in the solar interior. Once the field has emerged into the atmosphere, the strong current of the cylinder drives a Lorentz force that shears the emerged magnetic field along its polarity inversion line (PIL). As it expands, there is a plasma pressure deficit within the centre of the emerging region \citep{dmac09b}, which plasma drains into. This combination of shearing and compression, which occurs self-consistently in flux emergence, produces a new flux rope in the atmosphere. Depending on the configuration of the coronal magnetic field, reconnection between it and the emerging flux can allow the flux rope to escape. \citet{archontis12} have performed a parameter study for this setup.

\citet{dmac09c} perform a complementary simulation to that of \citet{archontis08}, replacing the initial cylindrical flux tube with a toroidal one. They demonstrate that multiple flux ropes (and hence eruptions) can be produced from the same emerging flux region. Recently, \citet{moreno13} have reported multiple eruptions in a flux emergence model for coronal hole jets. Their initial condition contains a cylindrical flux rope placed in the solar interior. It emerges into a constant magnetic field that is at an acute angle to the plane-parallel model atmosphere. Previously, in field-free or horizontal field atmospheres, it was reported that the cylindrical model only produces one atmospheric flux rope. The results of \citet{moreno13} suggest that CME flux rope production has as much to do with the dynamics and magnetic field of the background atmosphere as it does with the initial geometry of the emerging flux rope. 

In this work we revisit the simulation of \citet{dmac09c}. By studying the magnetic topology of the emerging flux region, we can identify the importance of magnetic reconnection during each stage of its evolution. We analyse both of the two flux ropes that are produced and discuss how they differ in topology. The paper is outlined as follows: Section 2 describes the basic model and numerical setup. Section 3 contains the analysis of the simulation results. Section 4 concludes the paper with a summary and discussions of further work with links to solar eruptions.

\section{Model description}
\subsection{Basic equations}
The 3D resistive and compressible MHD equations are solved using a Lagrangian remap scheme \citep{arber01}. In dimensionless form, these are
\[
\frac{{\rm D}\rho}{{\rm D} t} = -\rho\nabla\cdot\bmath{u},
\]
\[
\frac{{\rm D}\bmath{u}}{{\rm D} t} = -\frac{1}{\rho}\nabla p + \frac{1}{\rho}(\nabla\times\bmath{B})\times\bmath{B}+\frac{1}{\rho}\nabla\cdot\bmath{T}+\bmath{g},
\]
\[
\frac{{\rm D}\bmath{B}}{{\rm D} t} = (\bmath{B}\cdot\nabla)\bmath{u} - \bmath{B}(\nabla\cdot\bmath{u}) +\eta\nabla^2\bmath{B},
\]
\[
\frac{{\rm D}\varepsilon}{{\rm D} t} = -\frac{p}{\rho}\nabla\cdot\bmath{u} + \frac{1}{\rho}\eta|\bmath{j}|^2 + \frac{1}{\rho}Q_{\rm visc},
\]
\[
\nabla\cdot\bmath{B} = 0,
\]
with specific energy density
\[
\varepsilon = \frac{p}{(\gamma-1)\rho}.
\]
The basic variables are the density $\rho$, the pressure $p$, the magnetic induction $\bmath{B}$ (referred to as the magnetic field) and the velocity $\bmath{u}$. $\bmath{j}$ is the current density, $\bmath{g}$ is gravity (uniform in the $z$-direction) and $\gamma =5/3$ is the ratio of specific heats. The dimensionless temperature $T$ can be found from
\[
T = (\gamma-1)\varepsilon.
\]
We make the variables dimensionless against photospheric values, namely, pressure $p_{\rm ph} = 1.4\times 10^4$Pa; density $\rho_{\rm ph} = 2\times 10^{-4}$kg~m$^{-3}$; scale height $H_{\rm ph}=170$km;  surface gravity $g_{\rm ph} = 2.7\times 10^2$ms$^{-2}$; speed $u_{\rm ph} = 6.8$km~s$^{-1}$; time $t_{\rm ph} = 25$s; magnetic field strength $B_{\rm ph} = 1.3\times 10^3$G and temperature $T_{\rm ph} = 5.6\times 10^3$K. In the non-dimensionalization of the temperature we use a gas constant $\mathcal{R}=8.3\times 10^{3}$m$^2$s$^{-2}$K$^{-1}$ and a mean molecular weight $\tilde{\mu}=1$. $\eta$ is the resistivity and we take its value to be $10^{-3}$. The fluid viscosity tensor and the viscous contribution to the energy equation are respectively
\[
\bmath{T} = \mu\left(\nabla\bmath{u}+\nabla\bmath{u}^{\rm T}-\frac{2}{3}\bmath{I}\nabla\cdot\bmath{u}\right),
\]
\[
 Q_{\rm visc} = \bmath{T}:\frac{1}{2}(\nabla\bmath{u}+\nabla\bmath{u}^{\rm T}),
\]   
where $\bmath{I}$ is the identity tensor. We take $\mu = 10^{-5}$ and use this form of viscosity to aid stability. The code accurately
resolves shocks by using a combination of shock viscosity and
Van Leer flux limiters, which add heating terms to the
energy equation.

The equations are solved in a Cartesian computational box of
(non-dimensional) sizes, [-80, 80]$\times$[-80, 80]$\times$[-20, 85] in the
$x$, $y$ and $z$ directions respectively. The boundary conditions are
closed on the top and base of the box and periodic on the sides. The computational mesh contains 300$^3$ points.

\subsection{Initial conditions}
The initial idealized equilibrium atmosphere is given by prescribing the temperature profile
\[
T(z) = \left\{\begin{array}{cc}
1-\frac{\gamma-1}{\gamma}z, & z < 0, \\
1, & 0 \le z \le 10, \\
T_{\rm cor}^{((z-10)/10)}, & 10 < z < 20,\\
T_{\rm cor}, & z \ge 20,
\end{array}\right.
\]
where $T_{\rm cor} = 150$ is the initial coronal temperature. The solar interior is in the region $z<0$, the photosphere and chromosphere lie in $0\le z \le 10$, the transition region occupies $10 < z < 20$ and the corona is in $z \ge 20$. The other state variables, pressure and density, are found by solving the magnetohydrostatic equation
\[
\frac{\rm d}{{\rm d}z}\left(p+\frac{B_c^2(z)}{2}\right) = -\rho g,
\]
where $B_c(z)$ is a hyperbolic tangent profile, so that the field is uniform in the corona and rapidly declines to zero at the base of the transition region. The strength of the coronal field is taken to be 0.01 (0.13 G). The orientation
of the field is chosen so that it is almost antiparallel to the field
of the emerging flux tube when they meet. This is along the negative $x$--direction in this simulation.

The initial toroidal flux rope, that is placed in the solar interior, has the form

\[
B_x = B_{\theta}(r)\frac{s-s_0}{r},
\]

\[
B_y = -B_{\phi}(r)\frac{z-z_0}{s}-B_{\theta}(r)\frac{xy}{rs},
\]

\[
B_z = B_{\phi}(r)\frac{y}{s} - B_{\theta}(r)\frac{x(z-z_0)}{rs},
\]
with
\[
r^2 = x^2 + (s-s_0)^2, \quad s-s_0 = r\cos\theta, \quad x=r\sin\theta,
\]
and
\[
B_{\phi}=B_0e^{-r^2/r_0^2}, \quad B_{\theta} = \alpha r B_{\phi} = \alpha rB_0e^{-r^2/r_0^2}.
\]
This is derived from a regular expansion of a Grad-Shafranov equation \citep{dmac09b}. The axis of the flux tube is positioned along the $y$-axis. $s_0$ is the major axis of the tube and $r_0$ is the minor axis. $z_0$ is the base of the computational box. $\alpha$ is the initial twist and $B_0$ is the initial axial field strength. A study of how varying these parameters affects flux emergence is presented in \citet{dmac09b}. In this paper, we adopt the same parameter values as \citet{dmac09c}. These are $B_0 = 5$, $\alpha = 0.4$, $s_0 = 15$, $r_0 = 2.5$ and $z_0 = -25$. To initiate the experiment, the entire tube is made buoyant. That is, a density deficit relative to the background density is introduced.

\section{Analysis}

\subsection{Rise phase}
\begin{figure*}
 \includegraphics[scale=0.9]{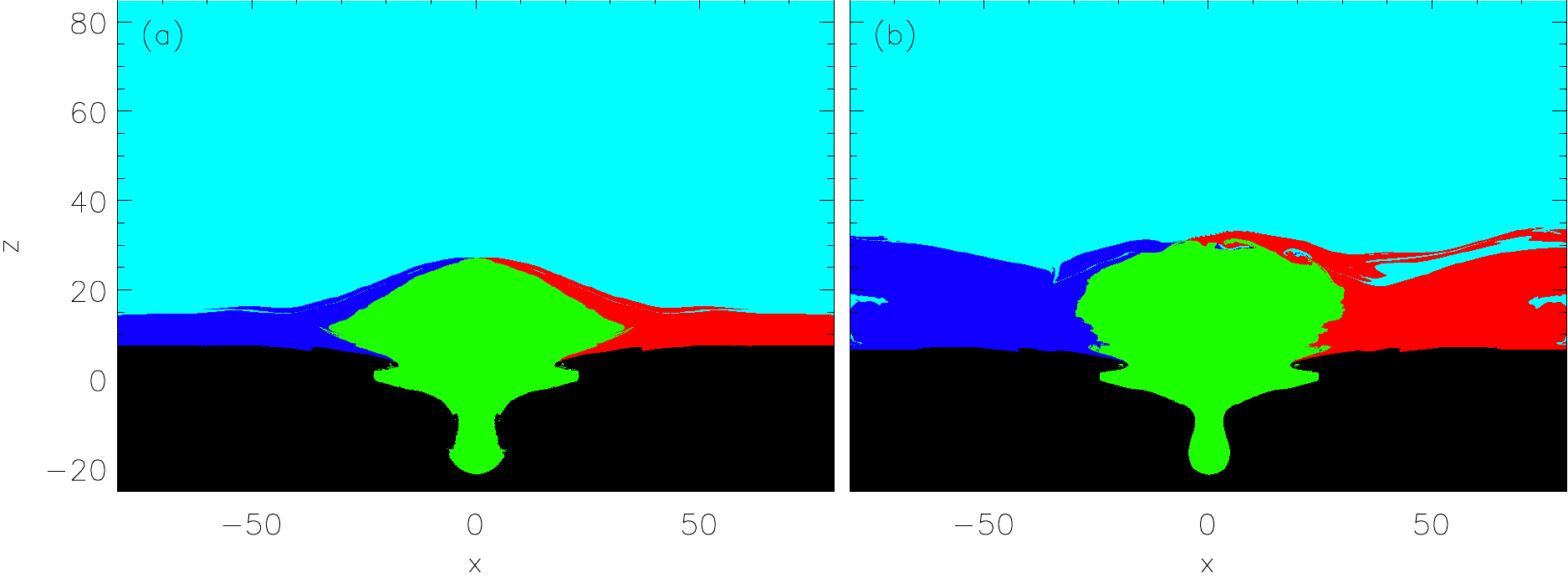}[h!]
 \caption{Connectivity maps of the magnetic field in the plane $y=0$ at (a) $t=63$ and (b) $t=74$. (a) shows reconnection at a single separator. (b) displays the formation of plasmoids. Key: cyan -- atmospheric flux, green -- emerging flux, blue -- flux from one side boundary in the $x$--direction conecting to one of the footpoints, red -- same as blue but for the other $x$--boundary and footpoint, black -- field free.}
\label{rise1}
\end{figure*}

The flux rope rises buoyantly to the photosphere and then expands, via the magnetic buoyancy instability, into the atmosphere. When the emerging magnetic field meets the atmospheric field, it is dynamically dominant with $\max_{z>0} |\bmath{B}|\approx 0.8$. As the emerging region pushes upwards into the overlying field, a current sheet forms between them and reconnection ensues. To visualize this, Figure \ref{rise1} displays connectivity maps for the magnetic field at (a) $t=63$ and (b) $t=74$ in the plane $y=0$. Magnetic field lines are traced throughout the domain and where they intersect the $y=0$ plane they are given a colour which depends on their connectivity. The green field lines are connected to both of the footpoints of the emerging flux region. The cyan field lines connect to the $x=-80$ and $x=80$ planes and represent the coronal magnetic field. Red field lines have been reconnected and connect from the $x=-80$ plane to one of the emerging flux rope's footpoints. Similarly, blue field lines connect from the $x=80$ plane to the other footpoint. Black regions are field-free.

\begin{figure}
\begin{center}
 \includegraphics[scale=0.30]{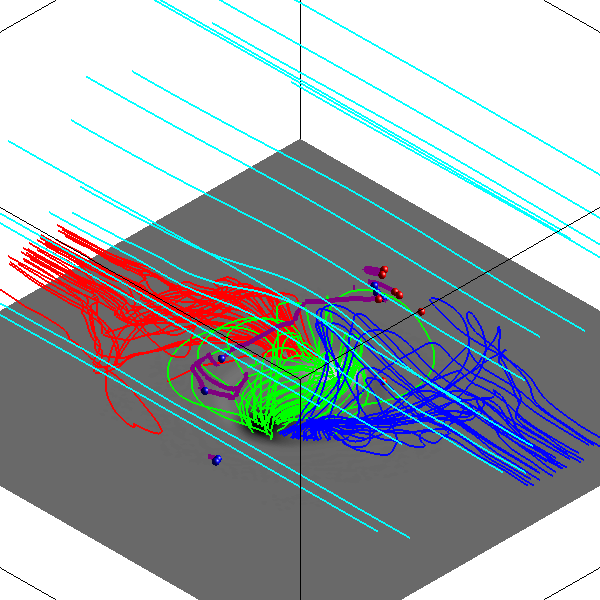}
\end{center}
 \caption{A plot of field lines at $t=63$ corresponding to the different flux regions. The field lines coloured in cyan, green, red and blue follow the same convention as in Figure 1. The purple field line that separates the four main regions is a separator. The separator connects to two clusters of nulls, shown as blue and red spheres. The slice shows $B_z$ at the base of the photosphere.}
\label{skeleton}
\end{figure}

As the (green) emerging flux system pushes into the (cyan) coronal flux, more coronal flux is reconnected and the blue and red regions grow in height. This is shown in Figure \ref{rise1} (a).
When the current sheet becomes thin enough, a more complicated phase of reconnection begins. Plasmoids form and are ejected out of the current sheet. Figure \ref{rise1} (b) shows this in the connectivity map at $t=74$.
In 2.5D flux emergence simulations \citep[e.g.][]{dmac09a,leake10}, similar behaviour is observed and is attributed to the tearing mode instability. In the standard 2D analysis of the tearing mode instability \citep[e.g.][]{paris84}, islands form in the current sheet and grow through reconnection at null points. This is not the case in 3D where more complex geometries exist. Instead of null point reconnection, the plasmoids are formed by separator reconnection \citep{parnell10a}. When reconnection first occurs between the emerging magnetic field and the coronal field, it takes place at one separator. This separator connects two clusters of null points on either side of the emerging flux region. These nulls exist in the model transition region and low corona. Figure \ref{skeleton} displays this separator, calculated from the magnetic skeleton \citep{haynes10}.
Similar behaviour has been found in other models of flux emergence \citep{maclean09,parnell10b}. The resulting magnetic topology of single separator reconnection (Figure \ref{rise1} (a)) is relatively straight forward. As the current sheet between the two flux domains becomes thinner, however, this separator undergoes several bifurcations and the number of separators increases \citep{parnell07}. This results in a much more complicated magnetic topology, as shown in Figure \ref{rise1} (b). In highly dynamic simulations such as this, it is diffucult to pin down whether or not the separator bifurcation is due to an instability or just plasma motions. It does share several similarities with the tearing mode instability in terms of the formation and ejection of plasmoids. We shall refer to the process as \emph{tearing reconnection} to distinguish it from other smoother reconnection events, such as when the two flux systems meet initially. We shall now discuss flux rope formation and the important role of reconnection.

\subsection{The first rope}
\begin{figure}
 \includegraphics[scale=0.60]{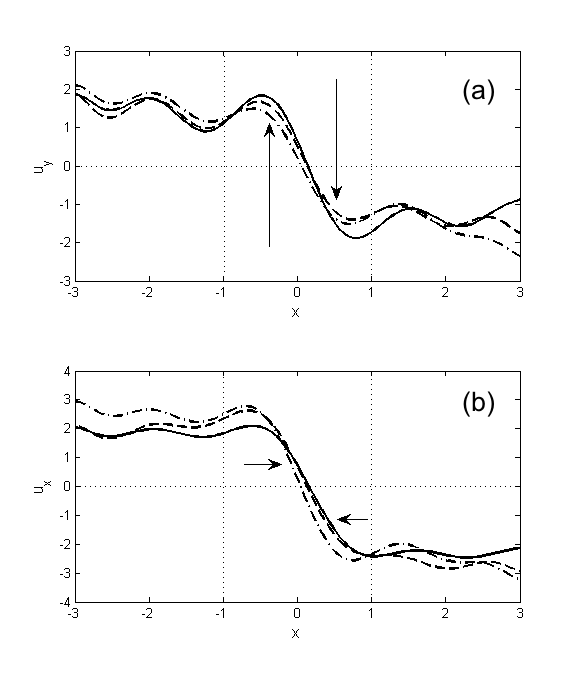}
 \caption{Flow profiles of (a) shear and (b) compression for the first rope in the plane $y=0$ at height $z=30$. The main focus is between the two vertial dotted lines near the PIL at $x\approx 0$. The arrows indicate the direction of the flow either side of the PIL. Key: solid -- $t=109$, double dash -- $t=110$, dot dash -- $t=111$.}
\label{u1}
\end{figure}

Informative models \citep{mackay10} have been successful in producing flux ropes from simple magnetic arcades. This is achieved through the imposition of shearing and compressional motions. In combination with these motions, flux ropes can be formed with reconnection \citep[e.g.][]{vanb89, kusano04}. A sheared and compressed arcade can also produce a `dipped field line' geomerty for prominences without reconnection \citep[e.g.][]{antiochos94}.
As mentioned in the Introduction, shearing and compression occurs self-consistently in the emergence of a twisted flux tube \citep{manchester04,dmac09b}. Figure \ref{u1} displays (a) the shear flows and (b) the compression flows at the PIL ($x\approx 0$) at times $t=$109, 110, 111. These are taken at a height $z=30$. (The reason for this will be made clear later by looking at the connectivity map.)   
From Figure \ref{u1} (a) the shear flows are approximately steady at the PIL over the time interval considered.
For the times shown, the compression speeds increase slightly with time but have a magnitude $|u_x|<3$ in the vicinity of the PIL.

\begin{figure}
 \includegraphics[scale=0.8]{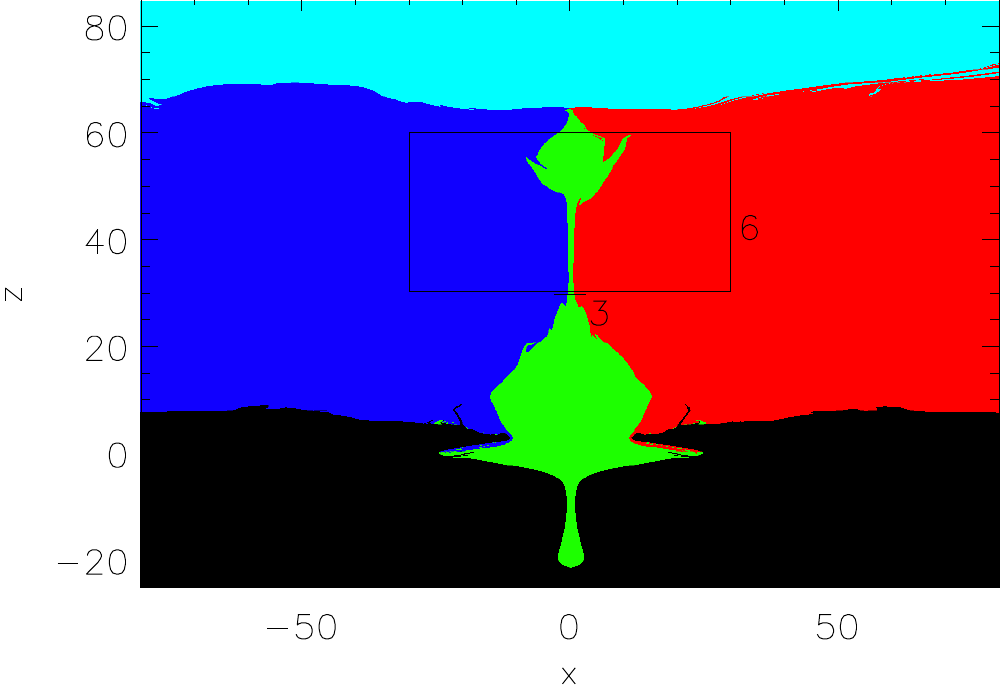}
 \caption{Connectivity map of the magnetic field in the plane $y=0$ at $t=111$. A rope-like structure forms at the top of the (green) emerging region. The horizontal black line labelled 3 shows the cut taken for the graphs in Figure 3. Similarly, the black box marks the region displayed in Figure 6.}
\label{fr_con}
\end{figure}

\begin{figure}
\begin{center}
 \includegraphics[scale=0.25]{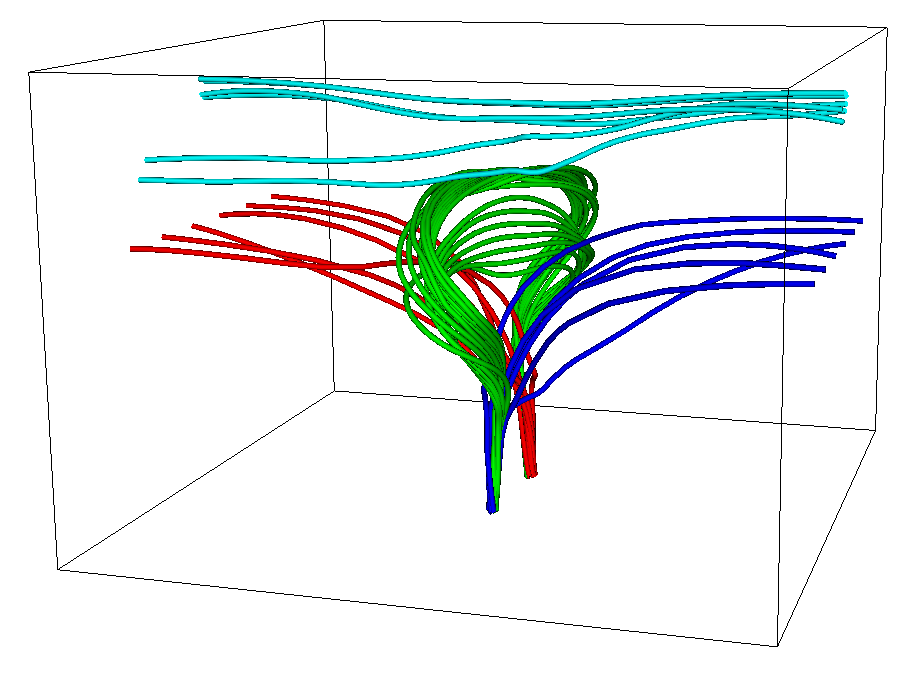}
\end{center}
 \caption{Magnetic field lines at $t=111$. The colour key is the same as the connectivity maps. A flux rope can be seen with a thin connective strip of emerging field beneath.}
\label{fr_fl}
\end{figure}

\begin{figure}
\begin{center}
 \includegraphics[scale=0.4]{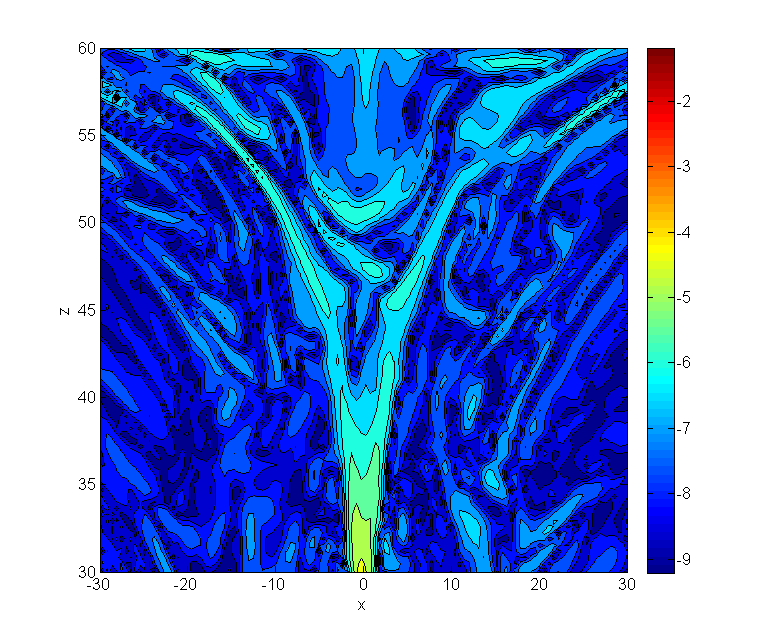}
\end{center}
 \caption{ A contour plot of $\log j_y$ in the $y=0$ plane at $t=111$. The plot highlights the top of the vertical current sheet and the position of the flux rope above it.}
\label{jy}
\end{figure}

\begin{figure}
\begin{center}
 \includegraphics[scale=0.25]{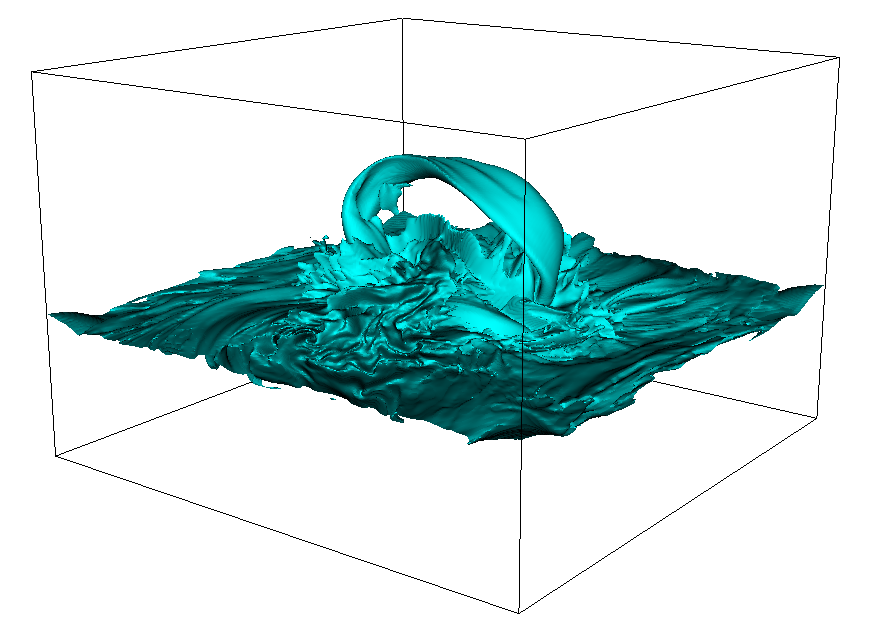}
\end{center}
 \caption{An isosurface of $\log\rho$ at $t=111$ displaying the geometry of the atmospheric flux rope.}
\label{rope}
\end{figure}

We shall now consider the `final state' of the first rope (its shape before it is dissipated at the top of the closed domain). Figure \ref{fr_con} displays the connectivity map at time $t=111$.
The emerging region has formed a rope-like structure at its height. Despite this formation, however, there is no splitting or fragmentation of the green region on the connectivity map. That is, although a flux rope forms in the atmosphere that carries dense plasma upwards, there exists a thin region of emerging magnetic field below the rope, starting from $z\approx 30$. Figure \ref{fr_fl} shows a 3D field line plot at $t=111$. (The colours of the field lines correspond to the same key as for the connectivity maps.)
Since the emerging flux domain has not split, there can be no magnetic separators threading underneath the flux rope.  Figure \ref{jy} displays a contour plot of $\log j_y$ in the $y=0$ plane at $t=111$. The vertical current sheet is clearly identifiable at $x=0$, the location of the sheared PIL. At $z\approx 40$ the current sheet bifurcates, forming a y-shape. Between the prongs of this y-shaped current sheet, at $z\approx 50$, there is an arc of enhanced $j_y$. This corresponds to the flux rope shown in Figures \ref{fr_con} and \ref{fr_fl} and is due to the compression of the magnetic field by dense plasma.    
Figure \ref{rope} displays the plasma density of the flux rope. The plasma follows the shape of the flux rope, a twisted sigmoidal loop, and closely resembles those obseved using \emph{SDO}. There is a clear distinction between the plasma in the rope and that in the connective strip.

In the analysis so far, it is plasma motion in the atmosphere that deforms the magnetic field to produce the flux rope. A sheared and compressed arcade forms during flux emergence and dense plasma collects at the top of this (in the y-shape of Figure \ref{jy}) to define the flux rope. That being said, our results do not preclude the possibility of 3D reconnection occuring at QSLs between the flux rope and the top of the vertical current sheet. Such (generalized) 3D reconnection could also help to shape the flux rope.

When the rope approaches the top (closed) boundary of the domain, it is dissipated away. The dense plasma carried up by the rope either drains down to the photosphere or is supported by the ambient magnetic field, remaining in the atmosphere.

\subsection{The second rope}

After the dissipation of the first rope, the emerging flux region continues to push upwards as it did before. This time, however, the geometry of the atmospheric magnetic field is different. During the formation of the first rope, the tension of the overlying coronal magnetic field was weakened by reconnection. This produced a `free path' for the emerging region and second flux rope to move into. Another feature of this is that the reconnected coronal field lines connect down to the footpoints, providing a downward path for draining plasma.

\begin{figure}
 \includegraphics[scale=0.55]{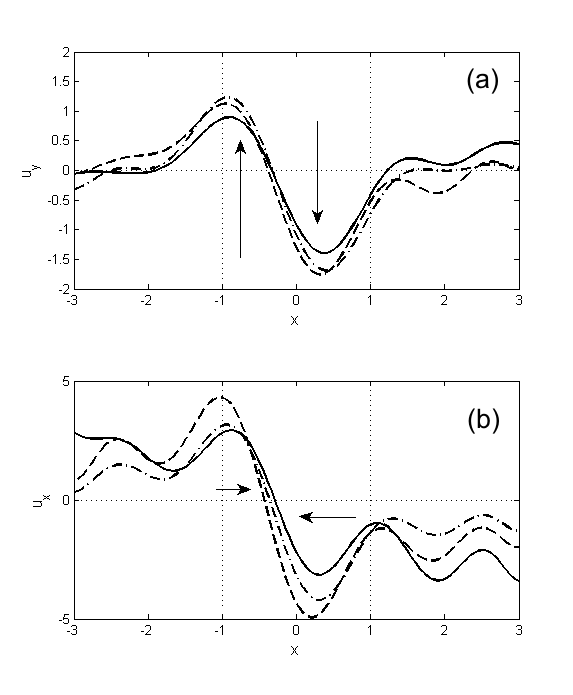}
 \caption{Flow profiles of (a) shear and (b) compression for the second rope in the plane $y=0$ at height $z=25$. The arrows indicate the direction of the flow either side of the PIL. Key: solid -- $t=131$, double dash -- $t=132$, dot dash -- $t=133$.}
\label{u2}
\end{figure}

As the new flux emerges, it is sheared along the PIL. Figure \ref{u2} (a) displays the shear profiles in the plane $y=0$, at height $z=25$ and times $t=$131, 132, 133.   
As during the formation of the first rope, the shearing speed maintains an approximately steady profile in time at the PIL. For the times shown there is a gentle acceleration. It should be noted that the magnitude of the shear flow during the formation of the second flux rope is slightly less than that during the formation of the first rope. However, the main point is that shearing continues during the formation of the second rope. 

\begin{figure}
\begin{center}
 \includegraphics[scale=0.25]{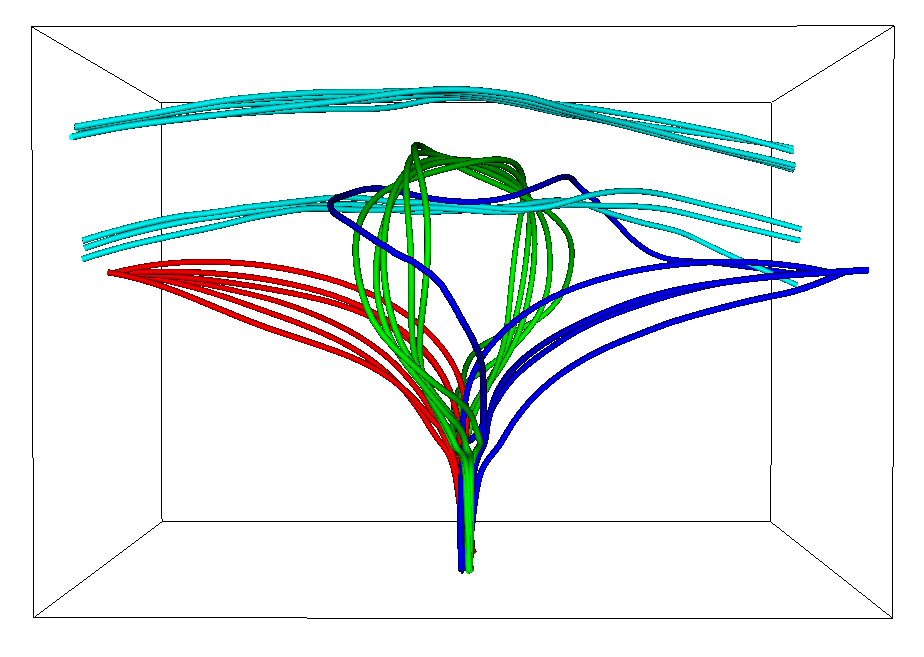}
\end{center} 
\caption{{Magnetic field lines at $t=136$. The colour key is the same as the connectivity map. The flux rope is now only connected to the photosphere at the footpoints. The magnetic field directly beneath the rope is connected elsewhere.}}
\label{sr_fl}
\end{figure}

As mentioned above, field lines from the corona connect down to the emerging region's footpoints at the start of the formation of the second rope. Draining plasma (brought up by emergence and left over from the first rope) drains down these reconnected field lines and results in stronger compressional flows near the PIL. Figure \ref{u2} (b) displays the compression profiles in the plane $y=0$, at height $z=25$ and times $t=$131, 132, 133.    
As with the first rope, the compression speeds at the PIL increase with time. However, the magnitudes are now greater than $|u_x|=3$ and, in places, are double the values of those for the first rope.  These enhanced speeds result in a different evolution for the second rope compared to the first.
\begin{figure*}
\includegraphics[scale=0.9]{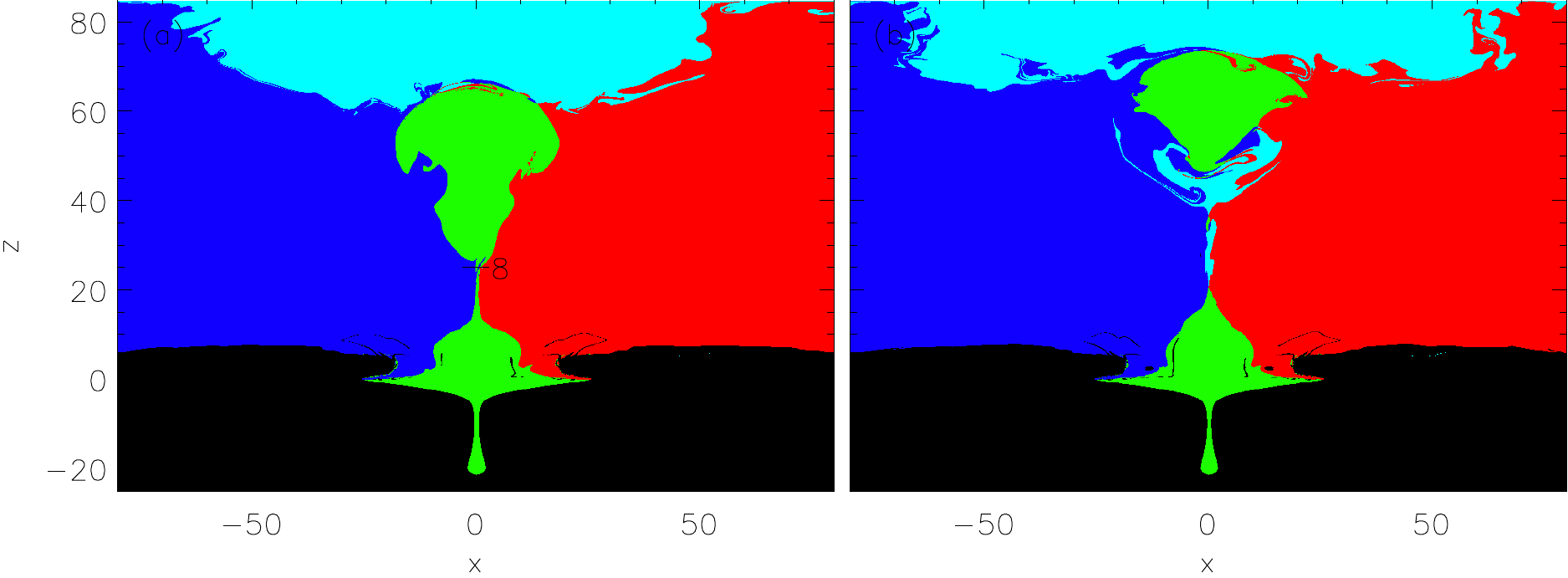}
 \caption{Connectivity maps in the plane $y=0$ at (a) $t=131$ and (b) $t=134$. These images show the connectivity before and after a significant change in topology due to tearing reconnection. The black horizontal line labelled 8 shows the cut taken for the graphs in Figure 8.}
\label{sr_con}
\end{figure*}
Figure \ref{sr_con} displays the connectivity maps at (a) $t=131$ and (b) $t=134$. At $t=131$ the shape of the (green) emerging region is similar to that during the evolution of the first rope. A rope-like region forms at the top with a thin region beneath connecting it to the flux at the photosphere. Due to the stronger compressional flows, however, tearing reconnection occurs. The map at $t=134$ displays that what was previously a thin green region has now split into islands.  In this simulation, points where all four colours (flux regions) meet are magnetic separators. In another context, this signature could also represent a quasi-separatrix layer (QSL) or a hyperbolic flux tube \citep[e.g.][]{titov09}. To visualize the effect of this tearing reconnection, Figure \ref{sr_fl} displays 3D magnetic field lines slightly later at $t=136$.
Unlike the first rope, magnetic field from the corona passes underneath the second rope. i.e. the second flux rope has a different magnetic topology to that of the first.

\section{Conclusions}
\subsection{Summary}
In this paper we present a simulation of solar flux emergence where two atmospheric flux ropes form self-consistently. Interestingly, their evolutions display important differences that are due to the dynamics in the atmosphere. When the emerging flux region first begins to expand in the atmosphere it reconnects with the overlying coronal field in two stages. To begin with, reconnection occurs at a single magnetic separator. This later splits into several separators and plasmoids are ejected from the current sheet between the emerging flux and the coronal field. Although these are both examples of separator reconnection, we refer to the latter as tearing reconnection as it is associated with the dynamic expulsion of plasma. 

Later, the emerging flux region produces an atmospheric flux rope. This rope is connected by a thin layer of magnetic field, over the PIL, down to the photosphere and below to the footpoints of the emerging region. The flux rope is created by smooth deformations in the atmosphere. One might describe this as an example of \emph{bodily} emergence. This term, however, does not reveal the full picture. The deformation of the expanded emerging field, by shearing and compression, plays an important role in moulding the geometry of the flux rope. Its structure is determined by the dynamics in the atmosphere, not just an upward expansion. The formation of an atmospheric flux rope primarily by deformation, rather than reconnection, has been inferred in other studies of dynamic flux emergence \citep{fan09,leake13b}. This is the first time, however, that the topology of the rope has been studied in detail. 

As the simulation domain is closed at the top, the first flux rope is dissipated away when it approaches close to it. Flux emergence continues, however, and magnetic field continues to push upwards from the photosphere. This time,    
the structure of the atmospheric magnetic field is different. Reconnection between the first rope and the coronal field has reduced the overlying tension above the emerging flux region. Also, field lines from the corona now connect down to the footpoints of the emerging region. Emergence proceeds as it did for the first rope. This time, however, due to stronger compressional flows, tearing reconnection occurs and a second flux rope forms that is only connected to the photosphere at its footpoints. i.e. it is a distinct loop. The reconnected coronal field lines that connect down to the footpoints allow plasma to drain efficiently. This plasma is a combination of that which is left from the first rope and that which is brought up by the second.

By analysing the magnetic topology, plasma flows and forces together, we have been able to gain a deeper understanding of the process of flux rope formation from magnetic flux emergence. 

\subsection{Discussion}   

The vicissitudes of flux rope formation from the same region may have interesting consequences. In this paper we have only considered formation, with eruption being left for further study. To investigate this, the upper boundary will have to be increased. Different profiles of the coronal magnetic field (decaying with height) will also have to be tested. This will help to determine what instabilities (if any) are responsible for the rise during eruption.  It also remains to be studied how the two different flux rope topologies will behave during an eruption. It may be the case that the first rope topology will not survive an eruption for two reasons. The first is that since the rope's plasma is not `trapped' inside a twisted magnetic rope, it will just drain away. The second is that if the vertical current sheet beneath rope becomes thin enough for a tearing instability to occur, the topology of the first rope might evolve into that of the second.

Another extension to the model would be to include the diffusive effects of Cowling resistivity in the chromosphere. As shown by \citet{leake13a}, this will affect the amount of sub-surface plasma raised to the corona during emergence.

The formation of the second rope has interesting connections to sympathetic eruptions - where magnetic activity in one region of the Sun has a causal link to an eruption in another region. Sympathetic eruptions have generated recent interest both observationally \citep{schrijver11,shen12} and theoretically \citep{torok11}. Although the draining plasma that helps create the second rope, in this paper, comes from one emerging region, the same effect is likely to occur if it flowed along field lines from another region. What this model suggests is that sympathetic interactions may be able to form atmospheric flux ropes as well as allow them to erupt \citep[as in][]{shen12}.  In effect, a rope with the topology of the first rope in our simulation could be converted into a rope with the topology of the second rope in our simulation via a sympathetic interaction. We shall pursue this line of enquiry in the near future.

\section*{Acknowledgments}

DM would like to thank the Royal Astronomical Society for helping to fund his visit to the National Astronomy Meeting 2013 in St Andrews. Part of this work is based on discussions between the authors during this meeting. The computational work for this paper was carried out on the joint STFC and SFC (SRIF) funded cluster at the University of St Andrews. We would like to thank the referee for helping to improve the clarity of this paper.

\label{lastpage}

\end{document}